\documentclass[%
 aip,
 amsmath,amssymb,
 reprint,%
]{revtex4-1}

\usepackage{graphicx}
\usepackage{dcolumn}
\usepackage{bm}

\usepackage[utf8]{inputenc}
\usepackage[T1]{fontenc}
\usepackage{mathptmx}
\usepackage{etoolbox}
\usepackage{xcolor}
\usepackage{subcaption}

\usepackage{comment}

\usepackage[hidelinks]{hyperref}

\makeatletter
\def\@email#1#2{%
 \endgroup
 \patchcmd{\titleblock@produce}
  {\frontmatter@RRAPformat}
  {\frontmatter@RRAPformat{\produce@RRAP{*#1\href{mailto:#2}{#2}}}\frontmatter@RRAPformat}
  {}{}
}%
\makeatother
\begin{document}

\preprint{AIP/123-QED}

\title{Machine Learning the order-disorder Jahn-Teller transition in LaMnO\textsubscript{3}}

\author{Lorenzo Celiberti}%
\thanks{These authors contributed equally to this work.}
\affiliation{University of Vienna, Faculty of Physics, Vienna, Austria}
\affiliation{University of Vienna, Vienna Doctoral School in Physics, Vienna, Austria}

\author{Alexander Ehrentraut}
\thanks{These authors contributed equally to this work.}
\affiliation{University of Vienna, Faculty of Physics, Vienna, Austria}

\author{Luca Leoni}
\affiliation{Department of Physics and Astronomy, University of Bologna, Bologna, Italy}

\author{Cesare Franchini}
\email[Corresponding author: ]{cesare.franchini@univie.ac.at}
\affiliation{University of Vienna, Faculty of Physics, Vienna, Austria}
\affiliation{Department of Physics and Astronomy, University of Bologna, Bologna, Italy}

\date{\today}

\begin{abstract}
We investigate the Jahn-Teller structural phase transition in LaMnO$_3$ at $T_{JT} \simeq 750$~K using molecular dynamics simulations based on machine-learning force fields trained on \emph{ab initio} data.
Analysis of the site–site correlation function of the distortions reveals that the transition is driven by the ordering of the $Q_2$ Jahn–Teller distortion of the MnO$_6$ octahedra, which acts as the order parameter and establishes the order–disorder nature of the transition. Dynamical local distortions are found to persist above $T_{JT}$.
Our results reproduce the experimental temperature dependence of both structural and phonon properties and highlight the presence of anharmonic effects at finite temperature. 
More broadly, the combined use of machine-learning molecular dynamics and velocity autocorrelation function analysis provides a robust framework for uncovering the microscopic mechanisms of structural phase transitions in correlated materials. In particular, this approach enables a clear distinction between order–disorder transitions and alternative mechanisms, such as displacive behavior, through the temperature evolution of vibrational properties.

\end{abstract}

\maketitle

\section{\label{sec:01}Introduction}

Phase transitions are ubiquitous in materials and encompass a wide variety of phenomena, ranging from classical transitions driven by thermal fluctuations, such as structural phase transitions, to quantum phase transitions controlled by interactions and quantum fluctuations, exemplified by superconductivity. From a theoretical perspective, the common language to decipher the evolution of a transition and understand the driving mechanism is statistical mechanics. To contribute to this Festschrift in honor of Christoph Dellago, we apply statistical-mechanical tools to interpret the temperature-driven order--disorder transition in the Jahn--Teller--active magnetic insulator LaMnO$_3$, a paradigmatic example of a correlated transition-metal system. The statistically relevant dataset is obtained using machine-learning force fields trained on accurate \emph{ab initio} simulations that explicitly include electron--electron correlations and magnetic interactions.

Transition metal oxides feature many of the most intriguing effects in condensed matter physics such as metal-insulator transition~\cite{Imada1998}, high-$T_c$ superconductivity~\cite{Bednorz1986}, colossal magnetoresistance (CMR)~\cite{Urushibara1995}, orbital physics~\cite{Tokura2000} and multiferroicity~\cite{Wang2009}.
The perovskite LaMnO$_3$ (LMO), the parent compound of colossal magnetoresistance manganites, is a prototypical transition-metal oxide in which several key aspects of correlated-electron physics coexist.
In LMO the Mn$^{3+}$ ion adopts a high-spin $d^4$ configuration, with the crystal field of the surrounding oxygen octahedron splitting the Mn $d$ levels into lower-energy $t^3_{2g}$ and higher-energy, doubly degenerate $e^1_g$ states. 
The partial occupation of the $e_g$ manifold renders this configuration unstable, and the resulting Jahn–Teller (JT) effect~\cite{Teller1937,Wollan1955,Kanamori1960} lifts the degeneracy by coupling the electronic states to specific lattice distortions of the MnO$_6$ octahedra, described by the $E_g$ vibrational modes $Q_2$ and $Q_3$. This electron-lattice coupling is central to the structural, electronic, and magnetic properties of LMO.

\begin{figure*}
\includegraphics[width=\linewidth]{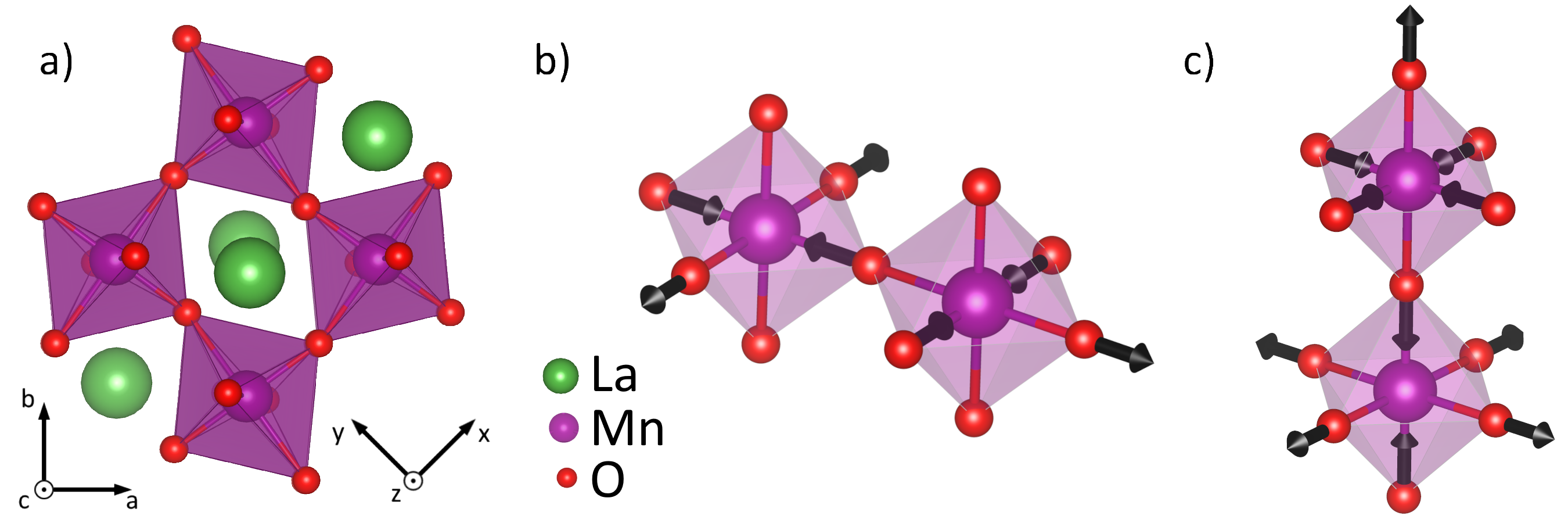}
\caption{a) {Crystal structure of orthorhombic LMO.}
The orthorhombic $Pbnm$ (a, b, c) and pseudocubic (x, y, z) axes are shown. 
{b) and c) distortion pattern of the cooperative JT modes $Q_2$ and $Q_3$ respectively.}
The VESTA software~\cite{Momma2011} was used for the creation of this figure.}
\label{fig:structure}
\end{figure*}

At low temperatures, LaMnO$_3$ is an orthorhombic ($Pbnm$) Mott-Hubbard insulator exhibiting cooperative Jahn-Teller distortions of the MnO$_6$ octahedra and long-range orbital order~\cite{Mihaly2004, Huang1997, Murakami1998, Sanchez2003, Wdowik2011} associated with the formation of two long, two medium, and two short Mn–O bonds, see Fig.~\ref{fig:structure}.
{In particular, the crystal field induced by the $Q_2$ distortions favors the occupation of different linear combinations of the Mn $e_g$ orbitals.
Depending on the sign of $Q_2$ orbitals with lobes pointing toward the $x$- or $y$-bond are occupied~\cite{Pavarini2010, Franchini_2012, Khomskii2021}.}
Below the N\'eel temperature $T_N \approx 140$~K, the coupling between orbital order and superexchange interactions stabilizes A-type antiferromagnetic order, with ferromagnetic alignment within the $ab$ planes and antiferromagnetic coupling along the $c$ axis~\cite{Wollan1955, Mihaly2004,Huang1997}.
The low temperature phase has been the subject of numerous ab-initio atomistic~\cite{He2012,Ergonenc2018,Varrassi2021,Lee2025} and effective Hamiltonian studies~\cite{Millis1996,Feiner1999,Ederer2007,Pavarini2010}
Upon heating above $T_N$, long-range magnetic order is lost and the system becomes paramagnetic, while remaining insulating. At higher temperatures, around $T_{JT}\simeq 750$~K, 
LMO showcases {an order-disorder} transition which results in a structural transition from orthorhombic to metrically (pseudo-cubic) cubic phase ~\cite{Rodriguez1998,Granado2000,Sanchez2003}. In this regime the cooperative Jahn-Teller distortion and static orbital order are suppressed, leading to a transition toward a more symmetric perovskite structure with dynamically fluctuating distortions. 
{Notably, the MnO$_6$ octahedra retain nonzero, disordered JT distortions up to 1150~K~\cite{Qiu2005,Sartbaeva2007,Pavarini2010}}.
This phase is particularly interesting because Sr or Ca doping transforms it into a ferromagnetic state exhibiting colossal magnetoresistance at low temperature~\cite{Qiu2005}.
At sufficiently high temperatures, LaMnO$_3$ approaches an almost undistorted perovskite phase, underscoring the strong interplay among lattice, charge, orbital, and spin degrees of freedom that governs its phase behavior~\cite{Pavarini2010,Thygesen2017}.

The microscopic origin of the unusual transition to the metrically cubic phase above $T_{JT}=750~K$
remains debated; however, there is broad consensus that it is closely related to the disordered distribution of Jahn–Teller–distorted octahedra~\cite{Murakami1998, Zhou1999, Qiu2005, Ahmed2009, Thygesen2017}. 
Standard first-principles atomistic calculations~\cite{Schmitt2020} and molecular dynamics simulations based on classical force fields~\cite{Qu2021} 
have helped to shed light on the Jahn-Teller transition; however, the small system sizes accessible to electronic-structure calculations and the absence of temperature-dependent electronic degrees of freedom in classical MD prevent an accurate description of this complex order-disorder electron–lattice phase transition.

The advent of machine-learned interatomic potentials (MLIPs), which enable large-scale and long-time molecular dynamics simulations with near first-principles accuracy, has opened new avenues for the modeling of complex phase transitions. Recent studies have demonstrated the capability of MLIPs to successfully capture the electron–lattice transition in LMO~\cite{Jansen2023,batnaran2025}.

Jansen \emph{et al}.~\cite{Jansen2023} reported that on-the-fly machine-learned force fields implemented in the Vienna Ab Initio Simulation Package (VASP~\cite{VASP1,VASP2}) predict an almost complete suppression of Jahn–Teller modes above 800~K, provided that an on-site Hubbard $U$, accounting for electron-correlation effects, is included in the training of the force fields. 
Based on this data set, Batnaran \emph{et al}.~\cite{batnaran2025} performed MD simulations using a parametrized ML interatomic potential based on NequIP~\cite{Batzner2022}, obtaining pair distribution functions and temperature-dependent octahedral distortion dynamics in agreement with experiment.

Here, building on the MLIP procedure adopted in Ref.~\onlinecite{Jansen2023}, we re-examine the JT transition in LMO by monitoring the temperature evolution of the Q$_2$ and Q$_3$ JT modes across the order–disorder transition.  
Our data describe well the progressive quenching of the JT modes with temperature and provide clear evidence for the persistence of \emph{dynamical} JT distortions at high temperatures, as extracted from the site-site correlation function.
{Furthermore, we calculated the phonon dispersion and identified the $A_{g}(1)$ mode having the same symmetry as the static distortion pattern in the ordered phase.
By Fourier transforming the velocity autocorrelation function of this mode, we obtained its temperature dependent spectral function, revealing strong anharmonic effects consistent with an order–disorder phase transition.}


\begin{figure*}
\includegraphics[width=0.9\linewidth]{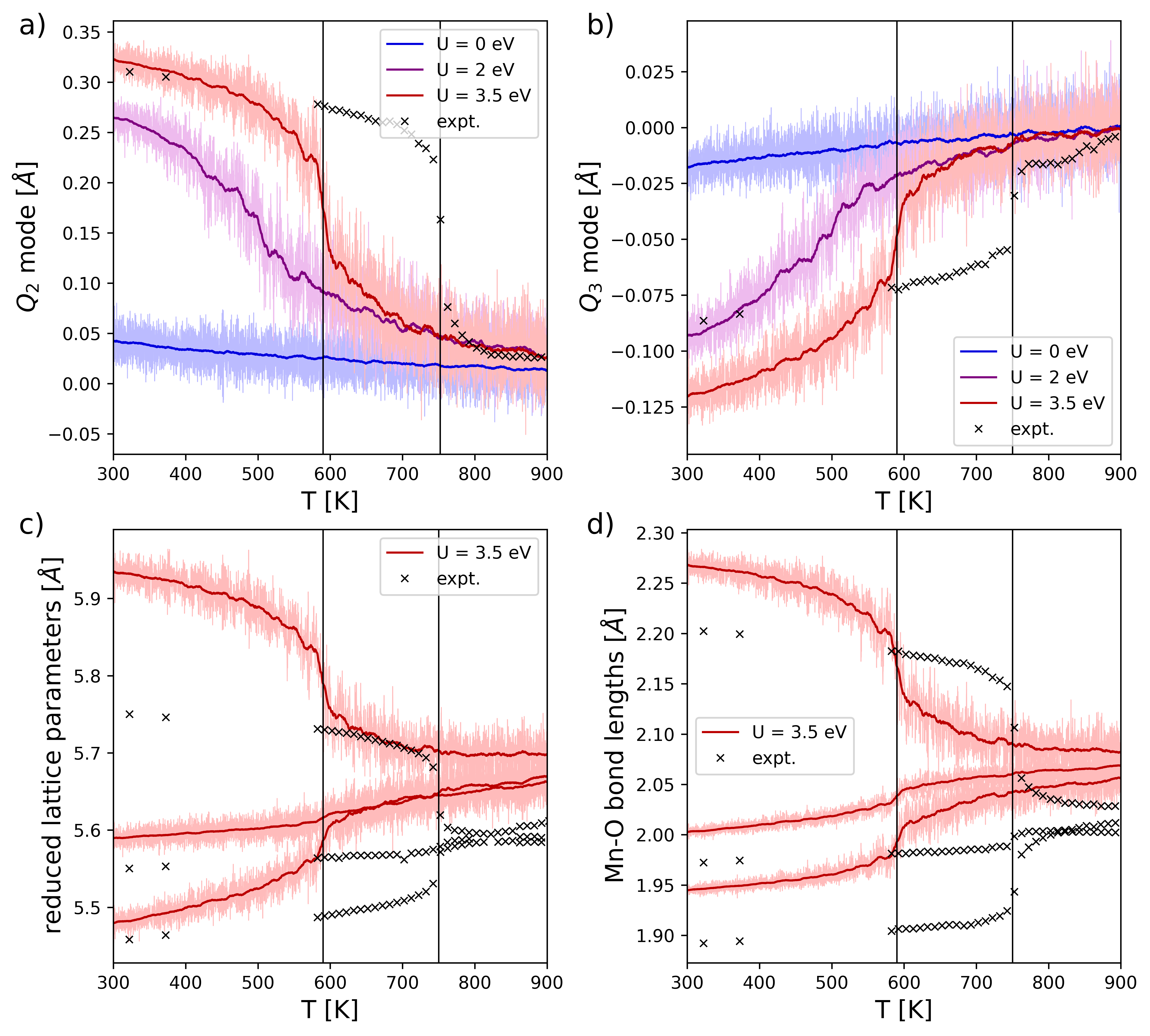}
\caption{\label{fig:tramp} a-b) Temperature evolution of the Q$_2$ and Q$_3$ JT modes during T-ramp obtained for different values of on-site $U$. c-d) Temperature evolution of reduced lattice parameters as defined in Ref. \onlinecite{Jansen2023} and Mn-O bond lengths during T-ramp obtained for on-site U = 3.5 eV. a-d) Oscillating signals (brighter colors) indicate the original time series while full lines (darker colors) represent a moving average of intervals of 400 configurations ($\sim$ 4000 time steps over $\sim$ 8 ps). Black crosses indicate experimental reference data  taken from Ref. \onlinecite{Thygesen2017}. Vertical bars indicate predicted (T$_{JT}^c$ = 590~K) and experimental (T$_{JT}$ = 750~K) transition temperatures.
}
\end{figure*}

\section{\label{sec:02} Methodology}

Our training and production simulations were conducted using the on-the-fly MLFF method implemented in the VASP package~\cite{VASP1,VASP2,Jinnouchi2019} within density functional theory (DFT) using the 
generalized Perdew-Burke-Ernzerhof (PBE)~\cite{Perdew1996} gradient approximation and including a local +$U$ correction according to the Dudarev approach~\cite{Dudarev}.

Following Ref.~\onlinecite{Jansen2023}, a $\sqrt{2}a \times \sqrt{2}b \times c$ supercell,
rotated by $45^\circ$ around the $c$ axis, containing eight formula units was employed for training the MLFF.  For the production runs, this supercell was
replicated three times along each lattice direction, resulting in a simulation cell comprising 216 formula units, corresponding to a total of 1080 atoms.
{A larger cell was also tested for finite size effects. We found no significant deviations (see Supplementary Materials).}

Force fields were trained for three values of the effective on-site interaction, $U_{\mathrm{eff}} = 0$, 2, and 3.5~eV. All training runs shared a $4 \times 4 \times 4$ \textit{k}-point grid and a plane-wave energy cutoff of
$E_{\mathrm{cut}} = 500$~eV. The magnetic moments were initialized in an A-type antiferromagnetic configuration, with local magnetic moments of $4\,\mu_{\mathrm{B}}$ on the Mn sites.

Molecular dynamics simulations were performed using time steps of 0.5~fs for the training runs, 2~fs for the NPT and NVT production runs and 0.4~fs for the NVE runs. The training simulations consisted of a linear temperature ramp from 600~K to 1100~K over a total of 125\,000 time steps, corresponding to a heating rate of 8~K/ps. The cutoff radii for the radial and angular descriptors were set to 6~\AA{} and 5~\AA{},
respectively.

Production simulations were carried out both under a linear temperature ramp from 100~K to 1100~K with a heating rate of 2~K/ps {tests for different heating rates are reported in the Supplementary Materials} and at fixed temperatures
of 100~K, 400~K, 600~K, 700~K, 800~K, and 900~K.  
A Langevin thermostat and the Parrinello-Rahman barostat were employed in the NPT and NVT simulations.

Analysis of the JT normal coordinates was done using the VanVleckCalculator~\cite{NagleCocco2024,NagleCocco2023}. 
The JT normal coordinates are defined as linear combinations of the Cartesian displacements $X_i$, $Y_i$, and $Z_i$ of the oxygen atoms relative to
their positions in an undistorted reference octahedron. In particular, we will focus our analysis on the relevant
Q$_2$ and Q$_3$ modes defined as (see Fig.\ref{fig:structure}b,c):

\begin{align}
Q_2 &= \frac{X_2 - X_5 - Y_3 + Y_6}{2},\\
Q_3 &= \frac{2Z_1 - 2Z_4 - X_2 + X_5 - Y_3 + Y_6}{2\sqrt{3}}. 
\end{align}

For phonon calculations we employed the finite difference method.
First, we relaxed a 4 formula unit cell, large enough to host the A-AFM and the checkerboard JT ordering, by using a $8\times 8\times 8$ $\Gamma$-centered k-points grid, with a plane-wave energy cutoff of 500~eV, a $10^{-6}$~eV energy tolerance, 0.001~eV/\AA force tolerance and $U=3.5$~eV.
We generated a $3\times 3\times 3$ supercell and displacements using Phonopy~\cite{Togo2023,Togo2023b}.
Energy and forces on the supercell were calculated only on the $\Gamma$ point.

\begin{figure*}
    \includegraphics[width=\linewidth]{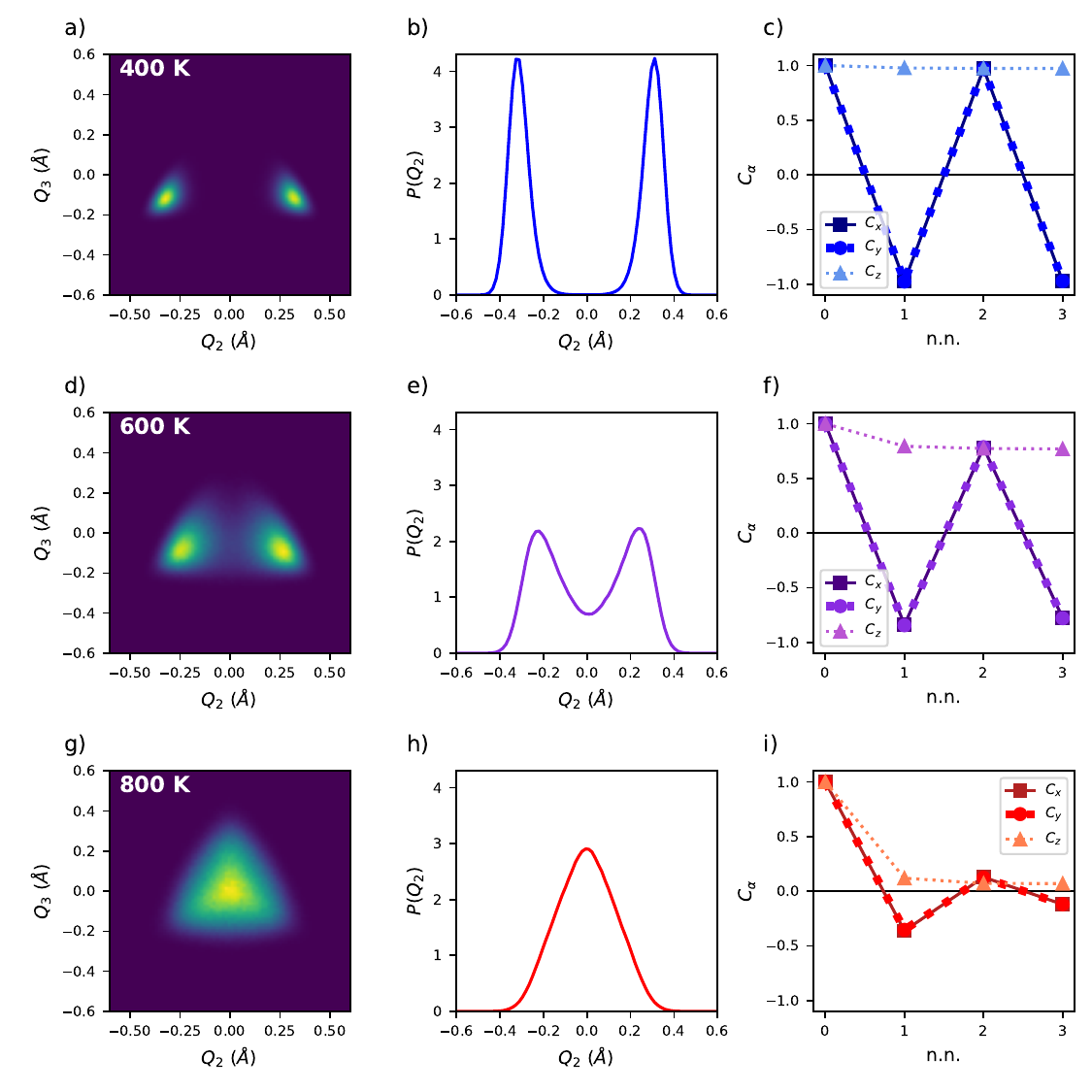}
    \caption{Statistical analysis of the Jahn-Teller vibrational modes $Q_2$ and $Q_3$ at three characteristic temperatures: $T = 400$~K (a--c), $600$~K (d--f), and $800$~K (g--i). Left: distribution of MnO$_6$ octahedra in the $(Q_2, Q_3)$ plane. Middle: probability distributions of the $Q_2$ mode. Right: site-site correlation functions.}
    \label{fig:static}
\end{figure*}

\section{\label{sec:03} Results and discussion}

We begin by validating our computational setup and MLIP predictions against available experimental data. Figure~\ref{fig:tramp} shows the temperature evolution of the relevant structural parameters.
Panels (a) and (b) highlight the crucial role of the Hubbard 
$U$ correction in capturing electron-lattice JT distortions in this correlated insulator~\cite{He2012, He2012b}. When 
$U$ is neglected, the Q$_2$ and  Q$_3$ modes are strongly underestimated and display only a weak temperature dependence, failing to show any signature of the structural transition. 
The inclusion of $U$ substantially improves the description, leading to a progressive suppression of the JT modes and octahedral rotation angles with increasing temperature, in agreement with Neutron-diffraction data~\cite{Rodriguez1998,Thygesen2017}. 
Therefore, all following reported results are obtained using $U$=3.5~eV. The predicted transition occurs around $T_{JT}^c \simeq 600$~K, which is lower than the experimentally observed transition at $T_{JT} = 750$~K (see vertical lines). 
{Increasing $U$ may improve the quantitative agreement between the calculated and experimental transition temperatures. 
However, it is well known that Dudarev’s correction with a single $U$ parameter, while yielding a satisfactory qualitative description of LMO, does not simultaneously reproduce quantitatively accurate electronic and lattice properties~\cite{Mellan2015}.}
More advanced exchange–correlation functionals, such as hybrid functionals~\cite{He2012b} or the random phase approximation~\cite{Jia2019, Verdi2023}, are known to yield quantitatively more accurate predictions, albeit at a significantly higher computational cost.

Further evidence supporting the ability of the MLIP to describe the transition is provided in Fig.~\ref{fig:tramp} (c,d), where we report the temperature dependence of the averaged lattice parameters and of the long, medium, and short Mn–O bond lengths, in comparison with X-ray and neutron scattering data~\cite{Thygesen2017}. Besides accounting for the low-T cooperative JT orbitally ordered state, the calculations reproduce the main features of the orthorhombic-to-pseudocubic transition, signaled by the progressive convergence of the lattice parameters toward similar pseudocubic values in the 570-620~K  temperature range. 

We note that the JT modes do not collapse to zero at the transition but instead persist at higher temperatures, in agreement with experimental observations, which is suggestive of an order-disorder dynamical transition.~\cite{Qiu2005,Sartbaeva2007}. 
These aspects are examined in detail in the remainder of this study.

Figure~\ref{fig:static} collects representative information at selected temperatures below (400~K), across (600~K), and above (800~K) the transition. At 400~K, the distribution of the octahedra within the ($Q_2$, $Q_3$) plane displays two bright spots (Fig.~\ref{fig:static}(a)), indicating a high probability for the octahedra to adopt these configurations, consistent with the C-type { ordering of the $Q_2$ modes.}
Across the transition at 600~K, these two features progressively merge into a broader region of the configurational space (Fig.~\ref{fig:static}(d)). 
At 800~K, the distribution becomes centered around the origin, signaling a strong suppression of long-range orbital order while retaining finite, dynamically fluctuating distortions.
Interestingly, the distribution shows only two peaks instead of the three expected from the tricorn potential of an isolated $e_g\otimes E_g$ JT site.
This deviation has been attributed by Popovic and Satpathy~\cite{Popovic2000} to the interaction between the JT modes in the crystal (cooperative JT effect).

The transition from a double to a single peak structure is transparently visualized by the corresponding probability distributions of Q$_2$ shown in the middle column.
These distributions are obtained by integrating over $Q_3$ the two dimensional distributions of Fig.~\ref{fig:static}(a,d,g).
The evolution from a two-peak structure to a broader distribution at high temperature is in good agreement with that of the correlations between quadrupole moments extracted from high-resolution neutron scattering data reported in Ref.~\onlinecite{Sartbaeva2007}. This behavior is not surprising, as the $E_g$ (and $T_{2g}$) JT distortions of an octahedron and quadrupolar moments both transform according to the same irreducible representations of $O_h$.

Next we study the auto- and cross-correlation functions of the $Q_2$ modes $C_{ij}(\tau)$  defined as
\begin{equation}
 C_{ij}(\tau) = \left< Q_{2,i}(t+\tau) Q_{2,j}(t) \right> 
 \label{eq:corr}
\end{equation}
where $i$ and $j$ are site indices and the average is taken over time and NVE MD trajectories~\cite{Lahnsteiner2022}. 
In particular, ten trajectories of 200~ps length where considered.
For $\tau=0$ we get the equal-time correlation function and, by averaging over site pairs with the same distance, we obtain the site-site correlation function resolved along the pseudocubic directions as
\begin{equation}
   C_\alpha(n) = \frac{1}{nN_{site}}\sum_{i=1}^{N_{site}} \sum_{|i-j|_\alpha = n} C_{ij}(0)
\end{equation}
where the second sum is restricted to the $n$-th nearest neighbor along the pseudocubic direction $\alpha = x,y,z$.
The results for 400~K, 600~K and 800~K are shown in Fig.~\ref{fig:static}(c,f,i) respectively.
For all three temperatures $C_x$ and $C_y$ coincide (squares and dots) and symmetrically oscillate between positive and negative values, whereas $C_z$ remains always positive.
This trend corresponds to the well known checkerboard Jahn-Teller distortion pattern of LMO~\cite{Pavarini2010,Rodriguez1998,Sanchez2003,Qiu2005,Sartbaeva2007} also depicted in Fig.~\ref{fig:structure}.
In particular, for $T<T_{JT}^c$ sites are perfectly (anti-)correlated independently from the distance, as expected in a long range order phase.
Slightly above the transition $T\gtrsim T_{JT}^c$, at 600~K, the correlation starts to decrease with octahedra separation but remain considerable even at the third nearest neighbor distance ($\sim 12$~\AA).
Above the transition the long range order is completely melted and the correlation remains sizable only at the first nearest neighbor.
At the same time, the $Q_2$ modes fluctuate at each octahedron around a small but non-zero value (see Fig.~\ref{fig:tramp}), thus confirming the order-disorder nature of the transition.

\begin{figure*}
    \centering
    \includegraphics[width=\linewidth]{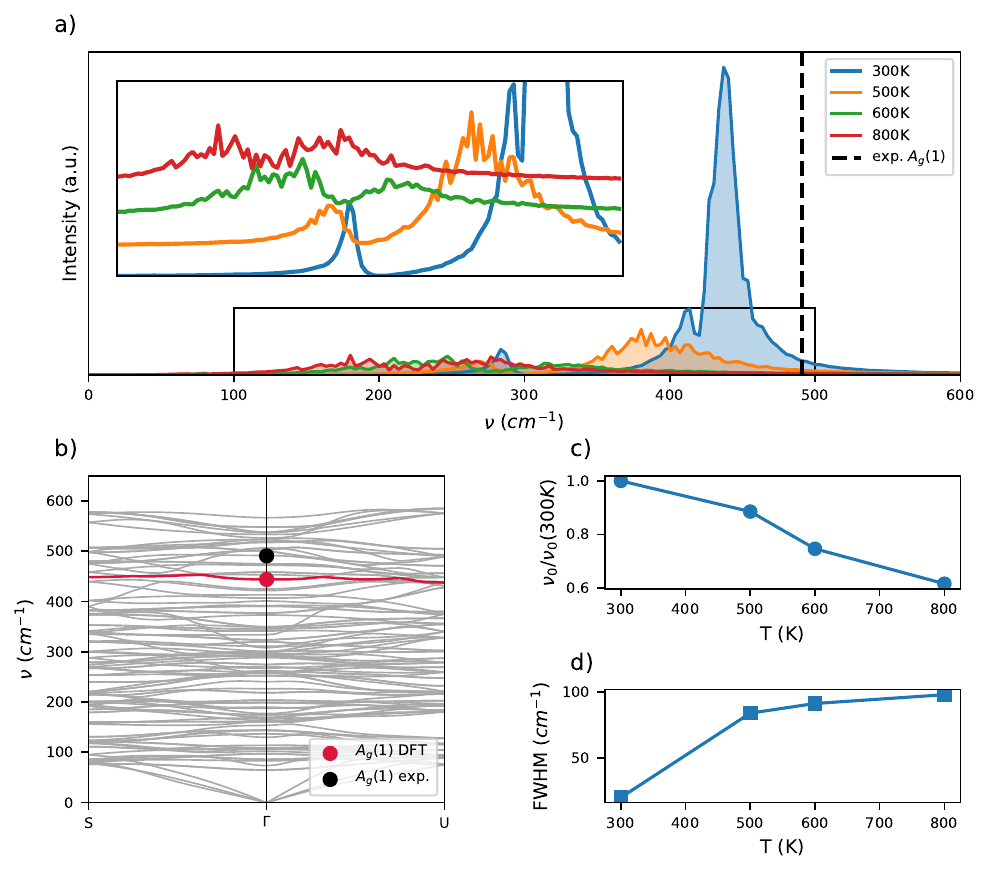}
    \caption{Spectral properties: a) Spectral function of the $A_g(1)$ mode as a function of temperature. b) phonon spectrum of the orthorhombic phase. The phonon band that at $\Gamma$ corresponds to the $A_g(1)$ mode is highlighted in red. c-d) 
   Parameters extracted from the Lorentzian fit of the mean peak: c) peak frequency $\nu_0$; d) full width at half maximum (FWHM).}

    \label{fig:dynamic}
\end{figure*}

In order to further characterize the transition, we move to the study of the velocity auto-correlation function (VACF). 
The Fourier transform of the velocity autocorrelation function (VACF), summed over all atoms, yields the phonon spectral function, while projecting the VACF onto the phonon eigenvectors allows one to obtain the spectral function of a specific mode.~\cite{Lahnsteiner2022}.
In particular, we projected the oxygen ions velocities along the pseudocubic directions with the right sign to obtain the $Q_2$ velocity $\dot{Q}_{2,i}$ of each octahedron.
By summing $\dot{Q}_{2,i}$ with a $(-1)^{i_x+i_y}$ phase we obtain the in-phase anti-stretching mode $A_g(1)$, which has the same distortion pattern of the static distortion in the ordered phase~\cite{Iliev1998, Iliev2006, Martin-Carron2001}. 
The spectral function of the $A_g(1)$ mode for 300~K, 500~K, 600~K and 800~K are reported in Fig.~\ref{fig:dynamic}(a). 

At 300~K a peak is observed at 438~cm$^{-1}$, slightly lower in frequency than the corresponding $A_g(1)$ phonon at 444~cm$^{-1}$ obtained from DFT (see Fig.~\ref{fig:dynamic}(b)).
The peaks frequency and width have been extracted with a Lorentzian fit, justified within the phonon quasi-particle picture~\cite{Sun2010,Lahnsteiner2022}.
The results are reported in Fig.~\ref{fig:dynamic}(c,d).
The peak at 300~K is relatively sharp and well described by a Lorentzian function, yet it exhibits a double-peak structure, indicating that anharmonic effects are already strong enough at 300 K to break the phonon quasiparticle picture.
The peak monotonically softens and broadens with increasing temperature, in good qualitative agreement with Raman measurements~\cite{Martin-Carron2001}.
However, we note that the $A_g(1)$ phonon frequency is underestimated by approximately 10\% with respect to the {value of 490~cm$^{-1}$ reported from experiments~\cite{Iliev1998,Iliev2006}.
Moreover, it exhibits a significantly stronger temperature-induced softening}. 
Consistently, the more rapid broadening is reflected in a faster increase of the FWHM compared to the experimental fit~\cite{Martin-Carron2001}, which may reflect an overestimation of anharmonic effects and the limitation of the DFT+U approach~\cite{Iliev1998,Iliev2006}.

Importantly, the progressive softening of the in-phase anti-stretching phonon mode across the transition is a clear fingerprint of its order–disorder character. This behaviour is fundamentally different from that observed in other types of phase transitions occurring in closely related transition-metal compounds. In incipient ferroelectrics such as KTaO$_3$~\cite{Ranalli2023} and SrTiO$_3$~\cite{Verdi2023}, the transition toward the ferroelectric state is predominantly displacive~\cite{schmidt2025,zhu2025}, with the soft anharmonic phonon driving the transition becoming progressively harder upon increasing temperature.

\section{CONCLUSION}


{In this work, we have clarified the nature of the structural phase transition in LaMnO$_3$ at $T_{JT} \simeq 750$ K through molecular dynamics simulations based on a machine-learning force field, complemented by analysis of distortion correlation and velocity autocorrelation functions. Our results demonstrate that the transition is driven by the ordering of the $Q_2$ Jahn–Teller mode of the MnO$_6$ octahedra.}

Our results reproduce the temperature dependence of the lattice parameters and provide compelling evidence that the symmetrized $Q_2$ distortion acts as the order parameter of the transition, in agreement with previous MLFF studies~\cite{Jansen2023,batnaran2025} and experimental observations~\cite{Thygesen2017,Sartbaeva2007,Popovic2000,Martin-Carron2001}.
The order-disorder nature of the transition is further established through site-site correlation functions, which reveal the disappearance of long-range order above the critical temperature and the emergence of small, spatially incoherent fluctuations of the Jahn-Teller $Q_2$ mode—hallmarks of a genuine order-to-disorder transition.

Additional insight is obtained from the analysis of phonon properties computed at the DFT+$U$ level for the magnetically ordered phase, together with the velocity autocorrelation function from molecular dynamics.
This combined approach enables a detailed characterization of the transition via the temperature evolution of the spectral function of the in-phase anti-stretching $A_g(1)$ phonon mode, in good agreement with Raman spectroscopy measurements~\cite{Martin-Carron2001}.
At the same time, the pronounced temperature dependence of this mode reveals strong anharmonic effects, which lead to a breakdown of the phonon quasiparticle picture already well below the transition temperature.

From a broader perspective, this work highlights the potential of combining machine-learning interatomic potentials with high-level \emph{ab initio} molecular dynamics as a general and powerful framework for the microscopic investigation of structural phase transitions in condensed matter systems. This approach enables direct access to finite-temperature lattice dynamics and local order parameters on length and time scales that are not accessible with either classical force fields or conventional \emph{ab initio} molecular dynamics alone. Within this framework, we identify the temperature evolution of the driving phonon mode and associated phonon properties as key fingerprints to discriminate between order-disorder and displacive transitions in transition metal perovskites, paving the way for systematic, predictive studies of anharmonic and electronically driven transitions in complex materials.

\section*{Supplementary Material}
The Supplementary Material provides additional details on the group–subgroup relationships of LaMnO$_3$, together with analyses of the training dataset, as well as heating rate and supercell size effects.

\begin{acknowledgments}
This research was funded by the Austrian Science Fund (FWF) 10.55776/F8100.
The computational results have been achieved using the Austrian Scientific Computing (ASC) infrastructure.
We gratefully acknowledge insightful discussions with Andrea Angeletti and Matthew Houtput.
\end{acknowledgments}

\section*{Data Availability Statement}

The data that support the findings of this study and the scripts used to extract them are available in the PHAIDRA repository of the University of Vienna at: \url{https://phaidra.univie.ac.at/o:2196925}.

\bibliography{biblio}

\end{document}



\title{Supplementary Materials: Machine Learning the order-disorder Jahn-Teller transition in LaMnO\textsubscript{3}}

\author{Lorenzo Celiberti}%
\thanks{These authors contributed equally to this work.}
\affiliation{University of Vienna, Faculty of Physics, Vienna, Austria}
\affiliation{University of Vienna, Vienna Doctoral School in Physics, Vienna, Austria}

\author{Alexander Ehrentraut}
\thanks{These authors contributed equally to this work.}
\affiliation{University of Vienna, Faculty of Physics, Vienna, Austria}

\author{Luca Leoni}
\affiliation{Department of Physics and Astronomy, University of Bologna, Bologna, Italy}

\author{Cesare Franchini}
\email[Corresponding author: ]{cesare.franchini@univie.ac.at}
\affiliation{University of Vienna, Faculty of Physics, Vienna, Austria}
\affiliation{Department of Physics and Astronomy, University of Bologna, Bologna, Italy}

\date{\today}

\maketitle
\renewcommand{\thefigure}{S\arabic{figure}}
\renewcommand{\thetable}{S\arabic{table}}

\section{Group-subgroup relationship}

\begin{figure}[h]
    \centering
    \includegraphics[width=0.6\linewidth]{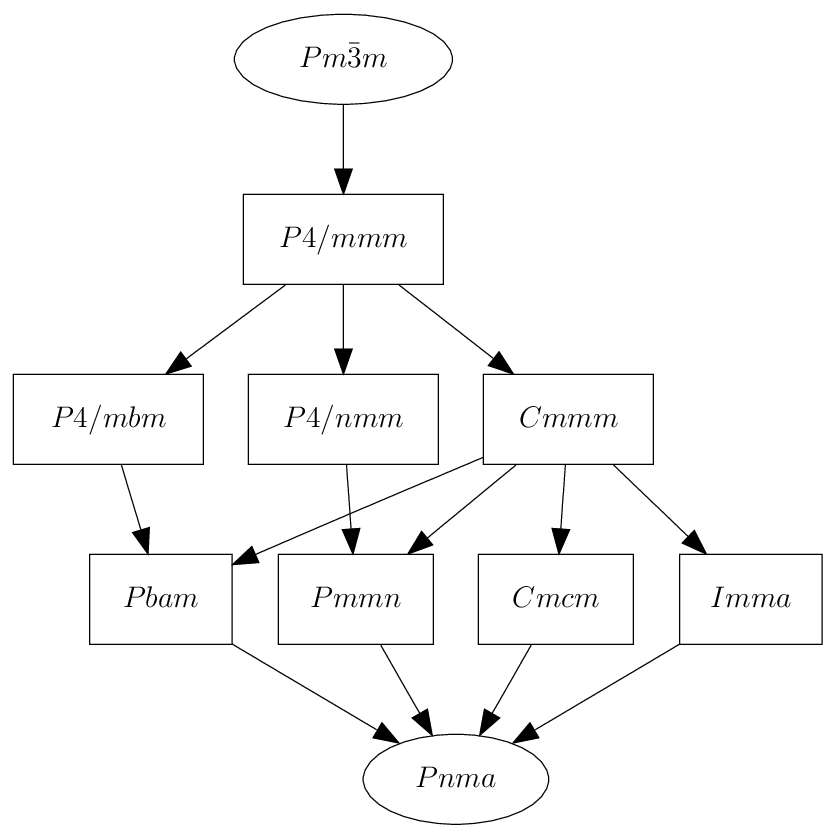}
    \caption{Group-subgroup graph for LaMnO$_3$.}
    \label{fig:group-subgroup}
\end{figure}

The $Pbnm$ (same space group as $Pnma$ with different labeling convention of the crystal axes) orthorhombic structure of LaMnO$_3$ can be obtained by distorting the standard $Pm\overline{3}m$ perovoskite structure.
In Fig.~\ref{fig:group-subgroup} we report the group-subgroup graph connecting $Pm\overline{3}m$ to $Pnma$ in LaMnO$_3$ obtained with SUBGROUPGRAPH tool~\cite{Ivantchev2000}.

The specific distortions of the $Pm\overline{3}m$ perovskite producing LaMnO$_3$ $Pnma$ structure can be obtained with the AMPLIMODES tool of the Bilbao Crystallography Server~\cite{Orobengoa2009, Perez-Mato2010}. 
We report them in Tab.~\ref{tab:distortions}.

\begin{table}[h]
    \centering
    \begin{tabular}{l|c|c|c|c|c}
         $k$-vector & Irrep. & Direction & Isotropy Subgroup & Dimension & Amplitude (\AA)  \\ \hline
         R=$(1/2, 1/2, 1/2)$ & $R_4^{-}$ & $(0, a, -a)$          & $Imma$ $(74)$    & 2 & 0.0623 \\
         R=$(1/2, 1/2, 1/2)$ & $R_5^{-}$ & $(0, a, a)$           & $Imma$ $(74)$    & 1 & 0.6886 \\
         X=$(0, 1/2, 0)$     & $X_5^{-}$ & $(0, 0, 0, -a, 0, 0)$ & $Cmcm$ $(63)$    & 2 & 0.3421 \\
         M=$(1/2, 1/2, 0)$   & $M_2^{+}$ & $(a, 0, 0)$           & $P4/mbm$ $(127)$ & 1 & 1.0155 \\
         M=$(1/2, 1/2, 0)$   & $M_3^{+}$ & $(a, 0, 0)$           & $P4/mbm$ $(127)$ & 1 & 0.3242 \\ \hline
    \end{tabular}
    \caption{Distortion modes of the standard perovskite cell giving rise to LaMnO$_3$ $Pnma$ unit cell.
             Amplitudes are normalized with respect to the primitive unit cell of the $Pm\overline{3}m$ structure.}
    \label{tab:distortions}
\end{table}

\newpage
\section{Training dataset analysis}

\begin{figure}[h]
    \centering
    \includegraphics[width=\linewidth]{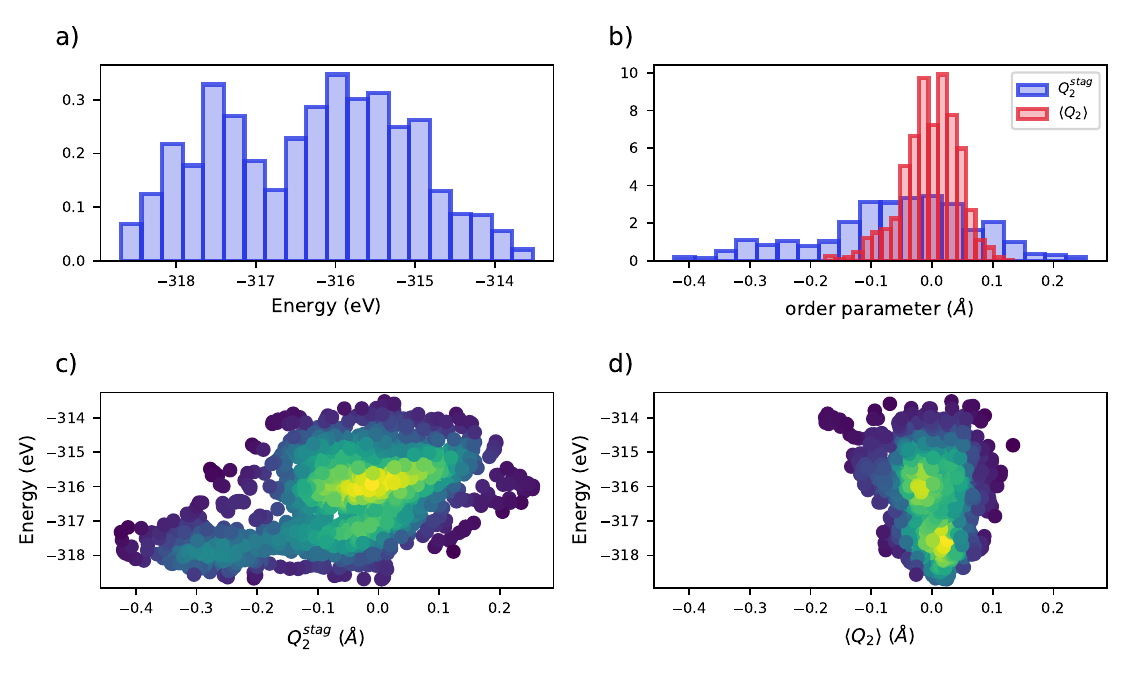}
    \caption{Training dataset analysis. 
    \textbf{a)} distribution of energies 
    \textbf{b)} distributions of $Q_2^{stag}$ (blue) and $\langle Q_2 \rangle$ (red) 
    \textbf{c)} distribution in the training structures in the $Q_2^{stag}$-Energy plane 
    \textbf{d)} distribution in the training structures in the $\langle Q_2 \rangle$-Energy plane.}
    \label{fig:dataset}
\end{figure}

In Fig.~\ref{fig:dataset} we analyze the training dataset contained in VASP's ML\_AB file. 
In particular, we report the distribution within the dataset of energies and values of the order parameter $Q_2^{stag}$  and average $Q_2$ distortion.
The dataset well represents both the ordered and disordered phases.

\newpage
\section{Heating rate and supercell effects}

\begin{figure}[h]
    \centering
    \includegraphics[width=\linewidth]{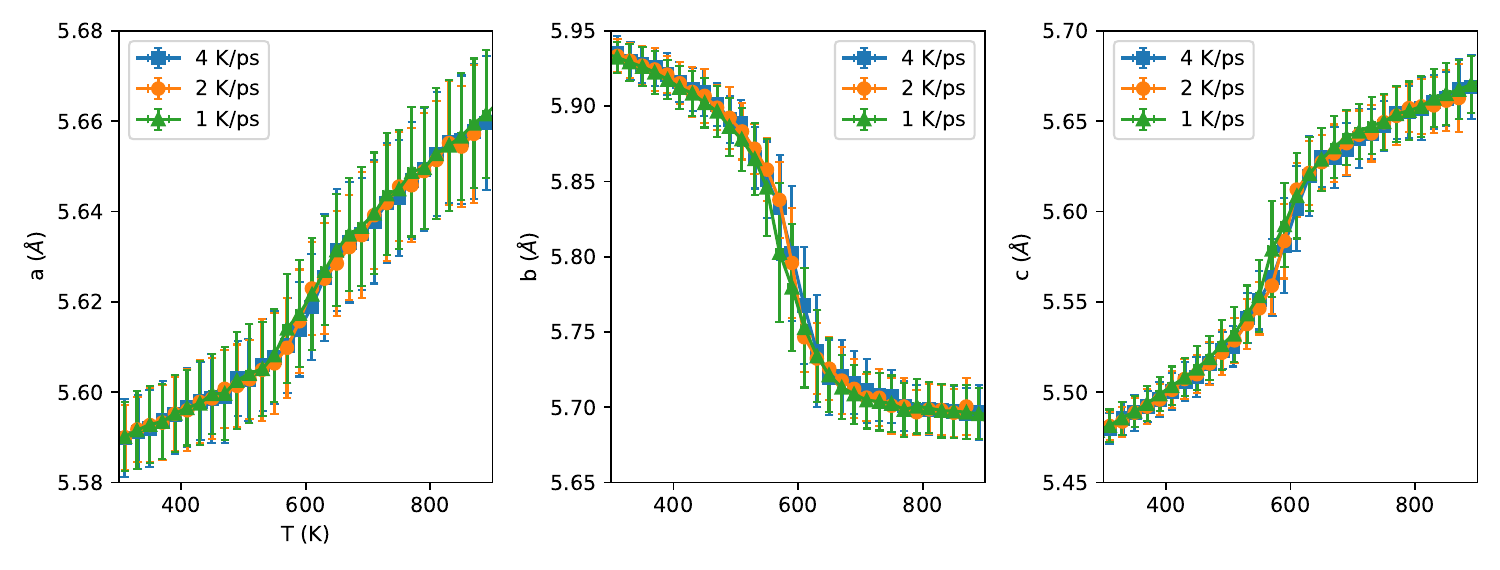}
    \caption{Reduced lattice constants as a function of temperature for different heating rates.}
    \label{fig:heating_rate}
\end{figure}

\begin{figure}[h]
    \centering
    \includegraphics[width=\linewidth]{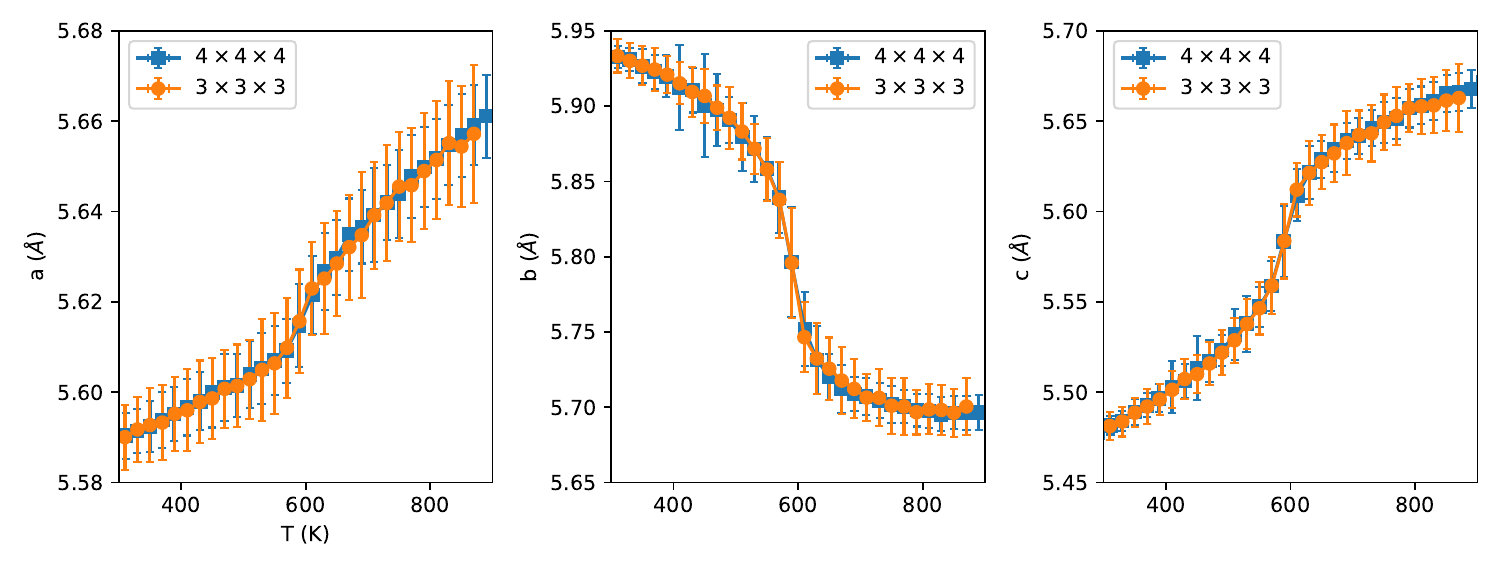}
    \caption{Reduced lattice constants as function of temperature for different supercell size.}
    \label{fig:supercell}
\end{figure}

We calculated the calculated the lattice parameters as function of temperature reported in Fig.~2(c) of the main text for different heating rates and for a larger $4\times 4\times 4$ supercell.
The results are reported in Fig.~\ref{fig:heating_rate} and Fig.~\ref{fig:supercell} respectively.
Error bars are estimated by performing a blocking analysis with a block dimension of 20 points.
We observe no significant deviation within the range of parameters explored.

\bibliography{supp_bib}